\documentclass[12pt]{iopart}

\begin{document}

\title{Symplectic space and orthogonal space of n qubits}
\author{Jian-Wei Xu}

\address{Key Laboratory for Radiation Physics and Technology of Ministry of
Education, Institute of Nuclear Science and Technology, Sichuan University,
Chengdu 610064, China} \ead{xxujianwei@yahoo.cn}

\begin{abstract}
In the Hilbert space of n qubits, we introduce the symplectic space (n odd)
and the orthogonal space (n even) via the spin-flip operator. Under this
mathematical structure we discuss some properties of n qubits, including
homomorphically mapping the local operations of n qubits into the symplectic
group or orthogonal group, and prove that the generalized ``magic basis'' is
just the bi-orthonormal basis (that is, the orthonormal basis of both
Hilbert space and the orthogonal space ). Finally, an example is given to
discuss the application in physics of this mathematical structure.
\\

\end{abstract}


\section{Introduction}

Quantum information theory is currently an active research field in physics.
Entanglement is one of its core concepts. Besides being of interest from a fundamental
point of view, entanglement also has many potential physical applications
such as quantum teleportation, quantum cryptography and quantum
algorithms \cite{1,2}.

We have already had a good understanding for the bipartite entanglement so
far. Entanglement of formation (EOF) and concurrence are the two widely used
measures of bipartite entanglement -- although the analytical expressions
are still limited to few special cases \cite{3}. However, the
situation of multipartite entanglement turns rather complicated. One basic
fact is, an n-body quantum state can be k-separable, i.e., it can be viewed
as the direct product of k parts, where $1\leq k\leq n$. To date, the
understanding of many-body entanglement is still very limited. This paper
will focus on the simplest many-body system, namely the n-qubit system,
which has received considerable attention \cite{5,6,7,8,9,10}.

How to operate a quantum system from one state to another? The so-called
LOCC (Local Operations and Classical Communication) and SLOCC (Stochastic
LOCC) are the natural and basic operations. We know that two (multi-body)
pure states can be transformed each other under LOCC if and only if they can
be transformed by local unitary transformations, and two (multi-body) pure
states can be transformed each other under SLOCC if and only if they can be
transformed by local transformations \cite{2}. So the mathematical
characterization of local operations $SU(n_1)\otimes SU(n_2)\otimes ...$ and
$SL(n_1)\otimes SL(n_2)\otimes ...$ becomes important in many aspects such
as entanglement classification and implementation of transformation of
many-body quantum system.

Uhlmann \cite{4} noted that the antilinear operators are intrinsically
nonlocal, so it is seemly natural to use antilinear operators to describe
the entanglement. Along this thought, some fruitful works \cite{5,6}
have been done. In \cite{5} the authors introduced the
pseudo-orthogonal basis to the Hilbert space of 2 qubits, and studied
the operators which preserve the pseudo inner product in the space spanned by the
pseudo-orthogonal basis. By this way, they realized the group
homomorphic mapping $SL(2)\otimes SL(2)$ to $SO(4,C)$. In other words,
for any local operation of $SL(2)\otimes SL(2)$, we can always find an
orthogonal matrix of $SO(4,C)$ to describe it. The present paper will
generalize this approach: in the Hilbert space $C_H^{2^n}$of n qubits, we
introduce the symplectic space $C_{HS}^{2^n}$ (n is odd) and the orthogonal
space $C_{HO}^{2^n}$ (n is even) via the spin-flip operator. We also define
the spin-flip operator on the linear operator spaces $L(C_{HS}^{2^n})$ and $%
L(C_{HO}^{2^n})$ which are spaces of all linear operators on $%
C_{HS}^{2^n},C_{HO}^{2^n}$, respectively. Under this mathematical structure
we discuss some properties of n qubits, including homomorphically mapping
the local operations of n qubits into the symplectic group or orthogonal
group, and prove that the generalized ``magic basis'' is just the
bi-orthonormal basis (that is, the orthonormal basis of both Hilbert space
and the orthogonal space ). We end by exhibiting an example to discuss the
application in physics of the mathematical structure in this paper.

Now I shortly call attention to some notations and rules, connected with
symplectic space and orthogonal space, concurrence and magic basis, to
prepare what follows below.

\subsection{Symplectic space and orthogonal space \cite{11}}

A vector space V ( we only limit to the finite dimensional vector spaces on
the complex field C) with a symmetric bilinear function $($ $,$ $)_O$ is
called an orthogonal space $V_O$, and $($ $,$ $)_O$ is called inner product
of the orthogonal space. The operators which preserve this inner product are
called orthogonal operators, all of which form orthogonal group. For any $%
\psi ,\varphi \in V_O,$ if $(\psi ,\varphi )_O=0,$ we say $\psi ,\varphi $
are orthogonal with respect to the inner product of $V_O$. A basis of $V_O$,
$\{\psi _i\}_{i=1}^n$, if satisfies $(\psi _i,\psi _j)_O=\delta _{ij}$, we
call it orthonormal basis with respect to the inner product of (regular)
orthogonal space. Orthogonal operators transform the orthonormal basis to
orthonormal basis.

Similarly, a vector space V with an antisymmetric bilinear function $($ $,$ $%
)_S$ is called a symplectic space $V_S$, and $($ $,$ $)_S$ is called inner
product of symplectic space. The operators which preserve this inner product
are called symplectic operators, all of which form symplectic group. For any
$\psi ,\varphi \in V_S$, if $(\psi ,\varphi )_S=0,$ we say $\psi ,\varphi $
are orthogonal with respect to the inner product of symplectic space. A
basis of $V_S$, $\{\psi _i\}_{i=1}^n$, if satisfies $(\psi _i,\psi
_j)_O=\varepsilon _{ij}$ (where  $\varepsilon
_{ij}=-\varepsilon _{ji}=1$ when j-i=1, otherwise $\varepsilon _{ij}=0$), we call it
orthonormal basis (or symplectic basis) with respect to the inner product of
(regular) symplectic space. Symplectic operators transform the orthonormal
basis to orthonormal basis. Obviously, the (regular) symplectic space must
be even dimensional.

\subsection{Concurrence and magic basis \cite{12,13}}

Wootters and Hill obtained the analytical expression of EOF for 2-qubit
states. During their research, they made use of the spin-flip operator (we use
symbol ``-'' to denote it), concurrence, and magic basis. For 1-qubit pure states, the spin-flip operator is defined as $|\overline{\psi }\rangle=\sigma _y|\psi ^{*}\rangle$, where * denotes complex conjugate, $\sigma _y$ is the Pauli operator. For 2-qubit pure states,
the spin-flip operator defined as $|\overline{\psi }\rangle=\sigma _y\otimes
\sigma _y|\psi ^{*}\rangle$, and it can be naturally extended for n-qubit pure
states, $|\overline{\psi }\rangle=\sigma _y^{\otimes n}|\psi ^{*}\rangle$. For 2-qubit
pure states, concurrence defined as $C(\psi )=|\langle\psi |\overline{\psi }\rangle|$.
For 2-qubit systems, given a local basis $\{|0\rangle,|1\rangle\}\otimes \{|0\rangle,|1\rangle\}$,
the so-called magic basis defined as
\begin{eqnarray}
|e_{1}\rangle=\frac{1}{2}(|00\rangle+|11\rangle),|e_2\rangle=\frac{i}{2}(|00\rangle-|11\rangle),\nonumber\\
|e_{3}\rangle=\frac{i}{2}(|01\rangle+|10\rangle),|e_4\rangle=\frac {1}{2}(|01\rangle-|10\rangle).
\label{mm}
\end{eqnarray}
An elegant result is, when we expand a 2-qubit pure state $|\psi \rangle$ in the
magic basis as $|\psi \rangle=\sum_i\psi _i|e_i\rangle$, the concurrence of $|\psi \rangle$
can then be expressed as
\begin{equation}
C(\psi)=|\sum_{i}\psi_{i}^{2}|.
\end{equation}
In this paper, we will generalize Eqs. (1) and (2) to
n-qubit (n even) cases.

\section{Symplectic space of single qubit}

In the two dimensional Hilbert space $C_H^2$ of 1-qubit system, for a given
local orthonormal (with respect to the Hilbert space) basis $\{|0\rangle,|1\rangle\}$, spin-flip operator is
defined on this basis as
\begin{equation}
|\overline{j}\rangle=(-1)^ji|j+1\rangle,j=0,1
\end{equation}
where $i$ is the imaginary unit, and $|1+1\rangle=|0\rangle$. For any two pure states $|
\psi\rangle ,|\varphi \rangle\in C_H^2,$ suppose $|\psi \rangle=\sum_j\psi _j|j\rangle,$ $|\varphi
\rangle=\sum_k\varphi _k|k\rangle,$ since spin-flip operator is antilinear operator, then
\begin{equation}
|\overline{\psi }\rangle=\sum_j\psi
_j^{*}|\overline{j}\rangle,  |\overline{\overline{\psi }}\rangle=-|\psi \rangle,
\end{equation}
\begin{equation}
\langle\overline{\psi }|\varphi \rangle=-i(\psi _0\varphi _1-\psi
_1\varphi _0).
\end{equation}
We define
\begin{equation}
(\psi ,\varphi )_S=\langle\overline{\psi }|\varphi
\rangle,
\end{equation}
Eq. (6) is antisymmetric bilinear with respect to $\psi
,\varphi $, so $C_H^2$ with Eq. (6) becomes a symplectic space, we
denote it $C_{HS}^2$.

It is easy to verify, $\{i|0\rangle,|1\rangle\}$
 is the orthonormal basis with respect to both the Hilbert space
and the sympletic space, we call it bi-orthonormal basis. Under this basis,
a symplectic operator S satisfies $S\left(\begin{array}{ccc}0&1\\ -1& 0\end{array}
\right)S^t=\left(\begin{array}{ccc}0&1\\ -1& 0\end{array}\right)$, where $t$ means matrix transpose. This condition is equivalent to detS=1. So, it follows that the special complex group SL(2,C) and symplectic group SP(1,C) are homomorphic,
\begin{equation}
SL(2,C)\sim SP(1,C).
\end{equation}
All the bi-orthonormal basis of $C_{HS}^2$ are
\begin{equation}
T\left(\begin{array}{ccc}i|0\rangle\\ |1\rangle\end{array}\right),T\in SU(2).
\end{equation}
For complex number $a$, we define $\overline{a}=a^{*}$, then for any two pure
states $|\psi \rangle,|\varphi \rangle\in C_H^2,$
\begin{equation}
\langle\overline{\psi }|\overline{\varphi
}\rangle=\overline{\langle\psi |\varphi \rangle}.
\end{equation}
Let $L(C_H^2)$ denote all linear operators on $C_H^2$. We also define the
spin-flip operator on $L(C_H^2)$, which is antilinear, and for any $|\psi
\rangle,|\varphi \rangle\in C_H^2,$
\begin{equation}
\overline{|\psi \rangle\langle\varphi |}=|\overline{\psi}\rangle\langle\overline{\varphi}|.
\end{equation}

Under the definitions above, for any $A,B\in L(C_H^2)$, $|\psi
\rangle,|\varphi \rangle\in C_H^2,$ $a,b\in C,$ we have the following properties
\begin{equation}
\overline{aA+bB}=a^{*}\overline{A}+b^{*}\overline{B},
\end{equation}
\begin{equation}
\overline{\overline{A}}=A,\overline{I}=I,
\end{equation}
\begin{equation}
\overline{A|\psi \rangle}=\overline{A}|\overline{\psi
}\rangle,
\end{equation}
\begin{equation}
\overline{AB}=\overline{A}$ $\overline{B},
\end{equation}
\begin{equation}
\overline{A^{\dagger }}=\overline{A}^{\dagger},
\end{equation}
\begin{equation}
\overline{A^{-1}}=\overline{A}^{-1}(A $ $ invertible).
\end{equation}
where $I$ is the identity operator of $L(C_H^2)$, $\dagger $ means Hermitian
adjoint, $-1$ the inverse.

\section{Orthogonal space and symplectic space of n qubits}

We use $C_H^{2^n}$ to denote the Hilbert space of n qubits. Now given a set
of local orthonormal bases $\{|0\rangle,|1\rangle\}$ of each qubit, denote them by $\{|j_1\rangle\},
\{|j_2\rangle\},...,\{|j_n\rangle\}$, or $\{|k_1\rangle\},\{|k_2\rangle\},...,\{|k_n\rangle\}$. For any $|\psi
\rangle,|\varphi \rangle\in C_H^{2^n}$, $|\psi \rangle=\sum_{j_1,j_2,...,j_n}\psi
_{j_1,j_2,...,j_n}|j_1,j_2,...,j_n\rangle,$ $|\varphi \rangle=\sum_{k_1,k_2,...,k_n}\varphi
_{j_1,j_2,...,j_n}|k_1,k_2,...,k_n\rangle,$ we have
\begin{equation}
|\overline{\psi }\rangle=\sum_{j_1,j_2,...,j_n}\psi
_{j_1,j_2,...,j_n}^{*}|\overline{j_1},\overline{j_2},...,
\overline{j_n}\rangle,
\end{equation}
\begin{equation}
\langle\overline{\psi }|\varphi
\rangle=\sum_{j_1,j_2,...,j_n}\psi _{j_1,j_2,...,j_n}\varphi
_{j_1+1,j_2+1,...,j_n+1}(-1)^{j_1+j_2+...+j_n}(-i)^n.
\end{equation}

We found that, when n is even, Eq. (18) is symmetric bilinear with
respect to $|\psi\rangle,|\varphi\rangle$, so $C_H^{2^n}$ becomes an orthogonal space, we
denote it by $C_{HO}^{2^n}$; meanwhile, when n is odd, Eq. (18) is
antisymmetric bilinear with respect to $|\psi \rangle,|\varphi \rangle$, hence $C_H^{2^n}
$ becomes symplectic space, we denote it by $C_{HS}^{2^n}$. Some direct
algebras will convince you that for n qubits cases also hold the properties
similar to Eqs. (9)-(16) of 1 qubit. In addition, for
local operations $A_1,A_2,...,A_n$, which operate on individual qubit
respectively, we have
\begin{equation}
\overline{A_1\otimes A_2\otimes ...\otimes
A_n}=\overline{A_1}\otimes \overline{A_2}\otimes ...\otimes
\overline{A_n}.
\end{equation}

\section{Orthogonal space $C_{HO}^{2^n}$ of n qubits (n even)}

\subsection{Bi-orthonormal bases of $C_{HO}^{2^n}$}

Suppose $\{|x_j\rangle\}_{j=1}^{2^n}$ is a bi-orthonormal basis of $C_{HO}^{2^n},$
that is, $\langle x_j|x_k\rangle=\delta _{jk},$ $\langle\overline{x_j}|x_k\rangle=\delta _{jk}.$
Fix one $|x_k\rangle,$ then for any $|x_j\rangle$ we have $\langle x_{j}-\overline{x_j}|x_k\rangle=0,$
hence $|\overline{x_j}\rangle=|x_j\rangle.$ We expand $|x_j\rangle$ in local basis $%
\{|0\rangle,|1\rangle\}^{\otimes n}$ (denote it by $\{\alpha _1\}\otimes \{\alpha
_2\}\otimes ...\otimes \{\alpha _n\}),$ then
\begin{eqnarray}
|x_j\rangle&=&\sum_{\alpha _1,\alpha _2,...,\alpha _n}x_{\alpha _1,\alpha
_2,...,\alpha _n}^j|\alpha _1,\alpha _2,...,\alpha _n\rangle,\nonumber\\
|\overline{x_j}\rangle&=&\sum_{\alpha _1,\alpha _2,...,\alpha _n}(x_{\alpha
_1,\alpha _2,...,\alpha _n}^j)^{*}(-1)^{\alpha _1+\alpha _2+...+\alpha
_n}i^n|\alpha _1+1,\alpha _2+1,...,\alpha _n+1\rangle.\nonumber
\end{eqnarray}
Since $|\overline{x_j}\rangle=|x_j\rangle,$ we obtain
\begin{equation}
x_{\alpha _1+1,\alpha _2+1,...,\alpha
_n+1}^j=(x_{\alpha _1,\alpha _2,...,\alpha _n}^j)^{*}(-1)^{\alpha _1+\alpha
_2+...+\alpha _n}i^n.
\end{equation}
$\langle x_j|x_k\rangle=\delta _{jk}$ then reads
\begin{equation}
	\sum_{\alpha _1,\alpha _2,...,\alpha _n}^{^{\prime
}}[x_{\alpha _1,\alpha _2,...,\alpha _n}^j(x_{\alpha _1,\alpha _2,...,\alpha
_n}^k)^{*}+(x_{\alpha _1,\alpha _2,...,\alpha _n}^j)^{*}x_{\alpha _1,\alpha
_2,...,\alpha _n}^k]=\delta _{jk},\label{(21.A)}
\end{equation}
where $\sum^{^{\prime }}$ means that in the sum indices $(\alpha _1,\alpha
_2,...,\alpha _n)$ and $(\alpha _1+1,\alpha _2+1,...,\alpha _n+1)$ only one
appears. If we let $x_{\alpha _1,\alpha _2,...,\alpha _n}^j=\frac 1{\sqrt{2}%
}(\lambda _{\alpha _1,\alpha _2,...,\alpha _n}^j+i\mu _{\alpha _1,\alpha
_2,...,\alpha _n}^j)$, where $\lambda _{\alpha _1,\alpha _2,...,\alpha
_n}^j,\mu _{\alpha _1,\alpha _2,...,\alpha _n}^j\in R,R$ is the real number
set, then Eq. (21) and $|x_j\rangle$ become
\begin{equation}
\sum_{\alpha _1,\alpha _2,...,\alpha _n}^{^{\prime
}}[\lambda _{\alpha _1,\alpha _2,...,\alpha _n}^j\lambda _{\alpha _1,\alpha
_2,...,\alpha _n}^k+\mu _{\alpha _1,\alpha _2,...,\alpha _n}^j\mu _{\alpha
_1,\alpha _2,...,\alpha _n}^k]=\delta _{jk},
\end{equation}
\begin{eqnarray}
|x_j \rangle=\frac1{\sqrt{2}}\sum_{\alpha _1,\alpha _2,...,\alpha _n}^{^{\prime }}[\lambda
_{\alpha _1,\alpha _2,...,\alpha _n}^j(|\alpha _1,\alpha _2,...,\alpha _n\rangle\nonumber\\
+(-1)^{\alpha _1+\alpha _2+...+\alpha _n}i^n|\alpha _1+1,\alpha _2+1,...,\alpha _n+1\rangle)\nonumber\\
+i\mu _{\alpha _1,\alpha_2,...,\alpha _n}^j(|\alpha _1,\alpha _2,...,\alpha _n\rangle\nonumber\\
-(-1)^{\alpha _1+\alpha _2+...+\alpha _n}i^n|\alpha _1+1,\alpha _2+1,...,\alpha _n+1 \rangle)].
\end{eqnarray}
We define the generalized magic basis as
\begin{eqnarray}
|e_{\alpha _1,\alpha _2,...,\alpha _n}^{+}\rangle=\frac 1{\sqrt{2}}[|\alpha _1,\alpha _2,...,\alpha _n\rangle\nonumber\\
+(-1)^{\alpha _1+\alpha _2+...+\alpha _n}i^n|\alpha _1+1,\alpha _2+1,...,\alpha
_n+1\rangle],\nonumber\\
|e_{\alpha _1,\alpha _2,...,\alpha _n}^{-}
\rangle=\frac i{\sqrt{2}}[|\alpha _1,\alpha _2,...,\alpha _n\rangle\nonumber\\
-(-1)^{\alpha _1+\alpha _2+...+\alpha _n}i^n|\alpha _1+1,\alpha _2+1,...,\alpha _n+1\rangle].
\end{eqnarray}
in Eq. (24), indices $(\alpha _1,\alpha _2,...,\alpha _n)$ and $(\alpha
_1+1,\alpha _2+1,...,\alpha _n+1)$ only one appears. We see that Eq. (24) is just the
generalization of Eq. (1). Combining Eqs. (24) and (23) we obtain
\\

\textbf{Theorem 1}:  $\{|x_j\rangle\}_{j=1}^{2^n}$ is a bi-orthonormal basis of $C_{HO}^{2^n}$ if and only if it can be expressed as
the generalized magic basis multiplied by a real orthogonal matrix.
\\

\subsection{Homomorphic mapping of local operations}

Suppose local operation $A_1\otimes A_2\otimes ...\otimes A_n$ preserves the
orthogonal inner product , i.e., it is an orthogonal operator, so for any $|\psi \rangle,|\varphi \rangle\in $ $C_H^{2^n}$,
$$\langle\overline{A_1\otimes A_2\otimes ...\otimes A_n\psi }|A_1\otimes A_2\otimes
...\otimes A_n\varphi \rangle=\langle\overline{\psi }|\varphi \rangle.$$
According to Eq. (19), we have
$(\overline{A_1})^{\dagger }A_1=I_1,(\overline{A_2})^{\dagger }A_2=I_2,...,(%
\overline{A_n})^{\dagger }A_n=I_n,$ where, $I_1$, $I_2,...,$ $I_n$ are the identities of individual qubit space.
It follows that $A_1,A_2,...,A_n$ are the symplectic operators of each $%
C_H^2 $. From Eqs. (7) and (14), we obtain
\\

\textbf{Theorem 2}: $SL(2)^{\otimes n}$ can be homomorphically mapped into $O(2^n,C)$
(n even)$.$
\\

A natural corollary of Theorem 2 is: if two states of $C_{HO}^{2^n}$
are connected by a operator which is not an orthogonal operator, then these two states cannot be
transformed by any local operation, i.e., they are not in the same SLOCC
class.

\section{Symplectic space $C_{HS}^{2^n}$ of n qubits (n odd)}

Suppose $\{|\alpha _1\rangle\},\{|\alpha _2\rangle\},...,\{|\alpha _n\rangle\}$, or $\{|\beta _1\rangle\},\{|\beta
_2\rangle\},...,\{|\beta _n\rangle\}$ are the local bi-orthonormal bases of individual
qubit, see Eq. (8), thus
$$\langle\alpha _1\alpha _2...\alpha _n|\beta _1\beta _2...\beta _n\rangle=\delta
_{\alpha _1\beta _1}\delta _{\alpha _2\beta _2}...\delta _{\alpha _n\beta _n},$$
$$\langle\overline{\alpha _1\alpha _2...\alpha _n}|\beta _1\beta _2...\beta
_n\rangle=\varepsilon _{\alpha _1\beta _1}\varepsilon _{\alpha _2\beta
_2}...\varepsilon _{\alpha _n\beta _n},$$
These imply that the $\{|\alpha _1\alpha _2...\alpha _n\rangle\}$ is exactly the
bi-orthonormal basis of $C_{HS}^{2^n}.$ Since unitary operators preserve the
inner product of Hilbert space, and symplectic operators preserve the inner
product of symplectic space, then we have
\\

\textbf{Theorem 3}: $\{|x_j\rangle\}_{j=1}^{2^n}$ is a bi-orthonormal basis of $C_{HS}^{2^n}$ if and only if it can be expressed as the basis $\{|\alpha _1\alpha _2...\alpha _n\rangle,each$ $|\alpha _i\rangle=i|0\rangle $ $ or $ $
 |1\rangle\}$ multiplied by a unitary-symplectic matrix (i.e., the matrix is both
unitary and symplectic).
\\

Similar to Theorem 2, we also have
\\

\textbf{Theorem 4}: $SL(2)^{\otimes n}$ can be homomorphically mapped into $%
SP(2^{n-1},C)$ (n odd).
\\

Similarly, a natural corollary of Theorem 4 is: if two states of $C_{HS}^{2^n}
$ are connected by a operator which is not a symplectic operator,
then these two states cannot be transformed by any local operation, i.e.,
they are not in the same SLOCC class.

\section{Quadratic form of $C_{HO}^{2^n}$}

We introduced the symplectic space and orthogonal space of n qubits, this
mathematical structure will bring conveniences for some problems. As an
example, we consider the quadratic form $\langle\overline{\psi }|\psi \rangle$ in $%
C_{HO}^{2^n}$ or $C_{HS}^{2^n}$. Obviously, $\langle\overline{\psi }|\psi \rangle$
vanishes in $C_{HS}^{2^n}$. However, in $C_{HO}^{2^n}$, $\langle\overline{\psi }%
|\psi \rangle$ may not vanish. In fact the absolute values
 $|\langle\overline{\psi }|\psi \rangle|$ form an entanglement measure of $|\psi \rangle$
\cite{14,15}, and it can be viewed as the formal
generalization of 3-tangle \cite{16}. Now we express $|\langle\overline{
\psi }|\psi \rangle|$ through the bi-orthonormal basis of $C_{HO}^{2^n}$, an
interesting result will occur. Assume we expand $|\psi \rangle$ in the
bi-orthonormal basis $\{|l\rangle\}_{l=1}^{2^n}$ as $|\psi \rangle=\sum_{l=1}^{2^n}\psi
_l|l\rangle,$ then we get
\begin{equation}
|\langle\overline{\psi }|\psi \rangle|=|\sum_{l=1}^{2^n}\psi
_l^2|.
\end{equation}
Eq. (25) is just the generalization of Eq. (2).

We now use Eq. (25) to describe the
maximally entangled states, i.e., $|\langle\overline{\psi }|\psi \rangle|=1$ for
normalized pure states $|\psi \rangle\in C_{HO}^{2^n}$. For this purpose, we first
make a geometrical interpretation of Eq. (25): we draw a polygonal line $OP_1P_2...P_n$ from O(0,0) on the complex plane such that $\overrightarrow{%
OP_1}=\psi _1^2,$ $\overrightarrow{P_1P_2}=\psi _2^2$,..., $\overrightarrow{%
P_{n-1}P_n}=\psi _n^2,$ then $|\langle\overline{\psi }|\psi \rangle|=|OP_n|$. Point $P_n$
cannot go beyond the unit circle, and the inequality $|\psi _l|^2\leq \frac
12(1+|\langle\overline{\psi }|\psi \rangle|)$ holds. When $P_n$ is on the unit circle, $%
|\psi \rangle$ is the maximally entangled state, in such case, the polygonal line $OP_1P_2...P_n$ becomes straight, with some
calculations, we obtain
\\

\textbf{Theorem 5}: Suppose $|\psi \rangle\in C_{HO}^{2^n},$ $\langle\psi |\psi \rangle=1.$ Under local
basis $\{|0\rangle,|1\rangle\}^{\otimes n}$ (denote it by
 $\{|\alpha _1\rangle\bigotimes|\alpha _2\rangle\bigotimes...\bigotimes|\alpha _n\rangle\}),$ $|\psi
\rangle=\sum_{\alpha _1,\alpha _2,...,\alpha _n}\psi _{\alpha _1,\alpha
_2,...,\alpha _n}|\alpha _1,\alpha _2,...,\alpha _n\rangle,$ and under
bi-orthonormal basis $\{|l\rangle\}_{l=1}^{2^n},$ $|\psi \rangle=\sum_{l=1}^{2^n}\psi
_l|l\rangle,$ then the following below are equivalent

(1). $|\langle\overline{\psi }|\psi \rangle|=1,$

(2). $|\psi \rangle=e^{i\theta }\sum_{l=1}^{2^n}\nu _l|l\rangle,$ where $\theta ,\nu
_l\in R,$ and $\sum_{l=1}^{2^n}v_l^2=1,$

(3). $|\psi \rangle=e^{i\theta }\sum_{\alpha _1,\alpha _2,...,\alpha _n}^{^{\prime
}}[\psi _{\alpha _1,\alpha _2,...,\alpha _n}|\alpha _1,\alpha _2,...,\alpha
_n\rangle$

$+(-1)^{\alpha _1+\alpha _2+...+\alpha _n}i^n\psi _{\alpha _1,\alpha
_2,...,\alpha _n}^{*}|\alpha _1+1,\alpha _2+1,...,\alpha _n+1\rangle],$

where $\theta \in R,$ and $\sum_{\alpha _1,\alpha _2,...,\alpha
_n}^{^{\prime }}|\psi _{\alpha _1,\alpha _2,...,\alpha _n}|^2=\frac 12.$

\section*{\textbf{Acknowledgements}}

This work was supported by National Natural Science Foundation of China
(Grant Nos. 10775101). The author thanks Li-Xiang Cen and Qing Hou for some
helpful discussions.

\end{document}